\documentclass[conference]{IEEEtran}
\renewcommand\IEEEkeywordsname{Keywords}
\IEEEoverridecommandlockouts
\usepackage{cite}
\usepackage{amsmath,amssymb,amsfonts}
\usepackage{algorithmic}
\usepackage{graphicx}
\usepackage{textcomp}
\usepackage{xcolor}
\usepackage{multirow}
\usepackage[nice]{units}
\usepackage{multirow}
\def\BibTeX{{\rm B\kern-.05em{\sc i\kern-.025em b}\kern-.08em
    T\kern-.1667em\lower.7ex\hbox{E}\kern-.125emX}}
\makeatletter
\def\endthebibliography{%
	\def\@noitemerr{\@latex@warning{Empty `thebibliography' environment}}%
	\endlist
}
\makeatother
\begin{document}

\title{Ecological Adaptive Cruise Control for City Buses based on Hybrid Model Predictive Control using PnG and Traffic Light Information\\
}

\author{\IEEEauthorblockN{Sai Krishna Chada$^{1}$, Jitin Mathew Thomas$^{1}$, Daniel G{\"o}rges$^{2}$, Achim Ebert$^{3}$, Roman Teutsch$^{4}$}
	\IEEEauthorblockA{\textit{$^{1}$Electromobility Research Group, University of Kaiserslautern, Germany} \\
		\textit{$^{2}$ German Research Center for Artificial Intelligence (DFKI), Kaiserslautern, Germany}\\
		\textit{$^{3}$Computer Graphics and HCI Group, University of Kaiserslautern, Germany}\\
		\textit{$^{4}$Institute for Mechanical and Automotive Engineering, University of Kaiserslautern, Germany}\\
		chada@eit.uni-kl.de, jthomas@rhrk.uni-kl.de, daniel.goerges@dfki.de, ebert@cs.uni-kl.de \& teutsch@mv.uni-kl.de}}
\maketitle
\begin{abstract}
This paper proposes an ecological adaptive cruise control (EACC) concept with the primary goal to minimize the fuel consumption in a city bus with an internal combustion engine (ICE). A hybrid model predictive control (HMPC) is implemented in this work to control both continuous and discrete-time variables. Moreover, a multi-objective optimization problem for EACC is formulated in time-domain as a mixed-integer quadratically constrained quadratic programming (MIQCQP) problem. The proposed HMPC-EACC performs robust vehicle-following while tracking a leading vehicle and plans fuel-efficient acceleration and deceleration maneuvers for the host vehicle. Additionally, it uses the signal phase and timing (SPaT) information to compute a green wave reference speed for the host vehicle to cross the signalized intersections at a green phase. Moreover, the proposed controller performs pulse and glide (PnG) to optimally control the engine ON and OFF states and save additional fuel. Furthermore, the performance of the proposed strategy is evaluated on a real-world driving profile and compared against a baseline controller from the literature. Finally, the influence of different prediction horizons on the fuel savings and computation times are studied. The results reveal significant reduction in fuel consumption with HMPC-EACC and demonstrate that the proposed controller is real-time capable.
\end{abstract}


\section{Introduction}
Local public transportation using city buses is regarded as an effective solution to mobilize people collectively in urban environments and reduce the need for private transportation, which is a major cause for traffic congestion and environment pollution. According to \cite{ACEA2019}, 98.3\% of the medium and heavy commercial vehicles in the European Union are diesel-powered, thus positioning the internal combustion engines (ICEs) as the dominant energy source in this segment. 
Although modern commercial vehicles are equipped with highly efficient ICEs which adhere to stringent emission norms, there is still a need to reduce the fuel consumption in the vehicle fleet, to lower the air pollution in the urban areas and to reduce the overall fleet operational costs as well. To address this concern, previous studies have investigated on varied aspects ranging from alternative fuels \cite{Imran2016}, electrified powertrain technologies \cite{Xu2013} and fuel-efficient routing approaches \cite{Suzuki2011}. Over the past decade, researchers have been actively investigating on advanced driver assistance systems (ADAS) that can promote energy savings in commercial vehicles and in turn help to reduce their CO2 emissions. Previous literature on energy-oriented ADAS in commercial vehicles focused on predictive/look-ahead cruise control \cite{Hellstrom2010} and eco-driving assistance systems \cite{Daun2013} which can achieve up to 12\% reductions in fuel consumption.
Recent studies point towards more advanced adaptive cruise control (ACC) concepts, in which a multi-objective optimal control problem (OCP) is solved with the aim to minimize the host vehicle's energy/fuel consumption, in addition to improving safety and driving comfort. One such concept is ecological adaptive cruise control (EACC), in which the host vehicle plans an optimal speed trajectory while tracking a preceding vehicle in a car-following scenario \cite{Li2009}, \cite{Jia2018}. Algorithms such as dynamic programming (DP)\cite{Hellstrom2010} and pontryagin’s minimum principle (PMP) \cite{Vajedi2016} have shown to yield a global optimum solution for the OCP, however were found to be computationally intensive and not suitable for the online implementation. To address this limitation, model predictive control (MPC) is being widely used to facilitate online-control and effectively deal with state and input constraints.

Fuel consumption in a conventional city bus is primarily influenced by the acceleration and deceleration maneuvers, resulting due to the lead vehicle behavior. To determine the interactions between a lead vehicle and host vehicle, previous works have explored car-following in passenger vehicles for highway driving 
and simplified urban scenarios. The influence due to other important factors in the urban environments such as traffic light (TL) signals, bus stops and stop signs have often been neglected in the literature. An effort was made in our previous work \cite{Chada2020} to propose a novel EACC concept to minimize the energy consumption in an electric vehicle (EV) by taking the advantage of signal phase and timing (SPaT) information from traffic signals. Recent advancements in the intelligent transportation systems (ITS) and vehicle-to-infrastructure (V2I) technology have made it possible to communicate between urban vehicles, signalized intersections and road infrastructure in real-time. 
One promising driving strategy to reduce the fuel consumption in a conventional vehicle is Pulse and Glide (PnG), in which the vehicle is accelerated to a higher speed than the desired speed in the pulse phase and then allowed to coast until a lower speed by turning the engine OFF in the glide phase. Such a strategy has been proven to be efficient in achieving fuel savings up to 20\% \cite{Li2012,Held2020}. To control the engine ON/OFF state, a binary variable must be incorporated into the optimization problem, thus making the hybrid model predictive control (HMPC) approach an ideal choice for the longitudinal control in conventional vehicles with ICE. To the best knowledge of the authors, EACC for city buses using HMPC is not yet addressed in the literature. 

In this work, the following contributions are made: (i) An ecological ACC strategy for a city bus with ICE based on HMPC is proposed, which is capable of reducing the host vehicle's fuel consumption, in addition to improving vehicle safety and driving comfort. For this purpose, a unique problem formulation is designed in time-domain to propel the host vehicle fuel-efficiently in the vehicle-following and freeway driving scenarios. 
(ii) The additional constraints due to TL signals, bus stops and regulatory speed limits present in an urban scenario are incorporated into the problem formulation. Moreover, the proposed HMPC-EACC uses the future information of the preceding vehicle velocities, route elevation and SPaT information to derive fuel-efficient velocity trajectories for the host vehicle. (iii) The optimization problem is formulated as a mixed-integer quadratically constrained quadratic programming (MIQCQP) problem and serves as an extension to the existing state-of-the-art EACC concepts. The performance of the proposed controller is evaluated against a problem formulation from the literature. (iv) Furthermore, a detailed comparison on the energy savings for different prediction horizons and the computation capability is presented in this work.  

The remainder of this paper is structured as follows. In Section \ref{sec:control_objectives}, the control objectives are introduced and a method for green-wave velocity calculation is presented. Later in Section \ref{sec:System_Dynamics}, the longitudinal dynamics of a city bus and the ICE scaling approach are elaborated. In Section \ref{sec:problem_formulation}, different problem formulations for EACC are compared and their respective properties are discussed in Section \ref{sec:results}. Furthermore, a detailed investigation on the computaton time for the proposed strategy is presented in Section \ref{sec:computation_time}. Finally, conclusions and hints on future work are given in Section \ref{sec:conclusions}.
\section{Control Objectives}
	\label{sec:control_objectives}
\subsection{Control objectives}
\label{control_objectives}
The primary objective of HMPC-EACC is to minimize the fuel consumption in a conventional city bus in the following scenarios: (i) While tracking a leading vehicle in a vehicle-following scenario by maintaining a relative distance not too large (to prevent cut-in scenarios) and not too small (to ensure safety). (ii) In a freeway scenario by tracking a reference green wave optimal speed (GWOS) devised for the host vehicle to pass through the signalized intersections at a green phase, thereby avoiding frequent stops at the TL signals. (iii) While approaching the bus stops and unavoidable red TL signals, the host vehicle must plan to stop fuel-efficiently. Furthermore, in such a scenario, the host vehicle must stop as close as possible to the stop line at TL signals or bus stops. (iv) In the aforementioned scenarios, the host vehicle must apply the PnG strategy and optimally turn the engine OFF when necessary. Besides these, other objectives such as vehicle safety and driving comfort must be ensured. 
\subsection{Green-wave velocity calculation}
\label{sec:green_wave_velcoity_calculation}
\begin{figure}[h]
	\centering
	\includegraphics[width=8.4cm]{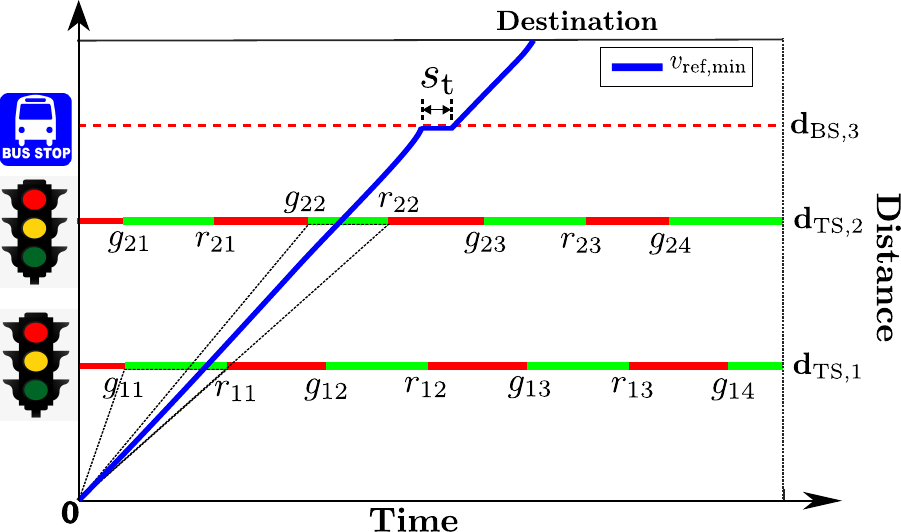}
	\caption{Space-time diagram showing an exemplary urban driving scenario}
	\label{fig:green_wave_demo}
\end{figure}

It is assumed in this work that the SPaT and bus stop information are communicated in real-time to the host vehicle. An examplary information is shown in Fig.~\ref{fig:green_wave_demo} using a space-time diagram. It can be noticed that the city bus driving route consists of two signalized intersections $\text{TS}_{\{1,2\}}$ and a bus stop $\text{BS}_{\{3\}}$. The approach used to calculate a green-wave velocity intersection is inspired from \cite{Asadi2011} and can be determined by,
	\begin{align}
[v_\text{ref,min} , v_\text{ref,max}] =	\left[\frac{d_{\text{TS},i\ }}{r_{ij\ }} ,\frac{d_{\text{TS},i\ }}{g_{ij\ }}\right] \cap [ v_\text{min} ,v_\text{max}] \ 
	\label{eqn:intersection}
	\end{align}	
where, $d_{\text{TS},i}$ represents the distance to the $i^{th}$ TL signal and $r_{ij}$ represents the start time of the $j^{th}$ red phase of the $i^{th}$ TL signal. Similarly, $g_{ij}$ represents the start time of the $j^{th}$ green phase of the $i^{th}$ TL signal. Furthermore, $v_\text{min}$ and $v_\text{max}$ are the minimum and maximum road speed limits respectively. It is to be noted that GWOS intersection using (\ref{eqn:intersection}) is calculated for the upcoming TL signals alone and bus stops are exempted because, it is considered in this work that the host vehicle makes a compulsory halt at all the bus stops. From the possible GWOS intersection range obtained using (\ref{eqn:intersection}), $v_\text{ref,min}$ is chosen as the new reference velocity, because travelling at lower velocities leads to better fuel-efficiency as compared to higher velocities. Therefore, the host vehicle controller tracks $v_\text{ref,min}$ to cross the TL signals at a green phase.

If a green-wave intersection range is not found in (\ref{eqn:intersection}), it means that stopping at the next TL signal is unavoidable. In such a case, the host vehicle is advised to travel at a driver's set speed until it reaches the upcoming TL signal. Similarly, if the host vehicle is approaching a bus stop, the phase is assumed to be red, so as to force the host vehicle to stop at bus stops for a designated time period called as bus dwell time (DT), usually for boarding or alighting the passengers, which is considered as $s_\text{t}=\unit[10]{s}$. Although the bus DT is not always constant in real scenarios, $s_\text{t}$ can easily be updated based on real measurements. Upon completion of this scenario, the GWOS is recalculated for the next TL signal.  

\section{System Modeling}
\label{sec:System_Dynamics}
\subsection{Vehicle dynamics}
The longitudinal dynamics of the host vehicle is given by
\begin{align}
a_\text{h} = \frac{dv_\text{h}}{dt} = \frac{1}{m_\text{eq}}(F_\text{t} - F_\text{b} - F_\text{res})
\label{eqn:Acceleration_1}
\end{align} 
	where $a_\text{h}$ and $v_\text{h}$ are the acceleration and velocity of the host vehicle, $m_\text{eq}$ is the equivalent mass which is the sum of vehicle weight, driver and cargo weight and rotational equivalent masses,  $F_\text{t}$ is the traction force and $F_\text{b}$ is the braking force. The resistance force is $F_\text{res} = F_\text{a} + F_\text{r} + F_\text{g}$, where $F_\text{a}$ is the aerodynamic resistance, $F_\text{r}$ is the rolling resistance and $F_\text{g}$ is the gradient resistance. The aerodynamic resistance force $F_\text{a}$ is calculated from the drag coefficient $c_\text{w}$, frontal cross-sectional area of the vehicle $A_\text{f}$, vehicle speed $v_\text{h}$ and density of the air $\rho$, as given in (\ref{eqn:aero_resistance}). The presence of the term $v^2$ makes the equation non-linear. $F_\text{a}$ can be linearized as 
\begin{align}
F_\text{a} = \frac{1}{2}\rho A_\text{f} c_\text{w} v_\text{h}^2 \approx \frac{1}{2}\rho A_\text{f} c_\text{w} (p_{1}v_\text{h}+p_\text{2}), 
\label{eqn:aero_resistance}
\end{align}
where the coefficients $p_\text{1}=27.711$ and $p_\text{2}=-168.459$ are identified through line approximation. Rolling resistance $F_\text{r}$ and gradient resistance $F_\text{g}$ are expressed as
\begin{subequations}
	\begin{flalign} 
	\label{eq:Resistance_forces_1}
	& \ \ \  F_\text{r} = c_\text{r}m_\text{v}g \cos\theta \\ \label{eq:Resistance_forces_3}
	&  \ \ \ \  F_\text{g} = m_\text{v}g \sin\theta
	\end{flalign}
	\label{eq:Resistance_forces}  
\end{subequations}
where $c_\text{r}$ is the rolling resistance coefficient, $m_\text{v}$ is the vehicle weight, $g$ is the gravitational acceleration and $\theta$ is the road elevation angle. 
The minimum safety distance $d_\text{s}$ is defined using the constant headway $h_\text{s}$ and the minimum safety distance at standstill $d_\text{min}$. To prevent frequent cut-in scenarios, the desired inter-vehicle distance $d_\text{c}$ is defined using the constant headway $h_\text{c}$ and maximum allowable distance $d_\text{max}$.
\begin{subequations} 
	\label{eq:SystemDyn_Cont}  
	\begin{align} 
	&	 d_\text{s} = d_\text{min} + h_\text{s} v_{\text{h},k} 
	\label{eqn:SystemDyn_Cont_4} \\
	&	d_\text{c} = d_\text{max} + h_\text{c} v_{\text{h},k} 
	\label{eqn:SystemDyn_Cont_5}
	\end{align} 
\end{subequations}
A reference engine map OM668 of a passenger vehicle is chosen from the ADVISOR library and is scaled to represent the city bus engine speed-torque characteristics using a simplified approach. Firstly, a torque scaling factor $f_\text{T,scale}$ is determined by dividing the maximum torque of the scaled engine $T_\text{e,max,new}$ with the maximum torque of the reference engine $T_\text{e,max,ref}$ as given in (\ref{eq:scaling}). By multiplying $f_\text{T,scale}$ with the torque values of the reference engine $T_\text{e,ref}$, the torque values for the scaled engine $T_\text{e,new}$ can be obtained. Similarly, a scaling factor for angular velocity $f_\text{w,scale}$ is derived by dividing the maximum angular velocity of the reference engine map $w_\text{e,max,ref}$ with the maximum angular velocity of the scaled engine $w_\text{e,max,new}$ as given in (\ref{eq:scaling_1}). Moreover, the new scaled angular velocities $w_\text{e,new}$ can be calculated by dividing the reference angular velocities $w_\text{e,ref}$ with the scaling factor $f_\text{w,scale}$.
\begin{subequations}
	\begin{flalign} 
	\label{eq:scaling}
	& f_\text{T,scale} =\frac{T_\text{e,max,new}}{T_\text{e,max,ref}} \ \Longrightarrow \ T_\text{e,new}=f_\text{T,scale} T_\text{e,ref}\\ 	\label{eq:scaling_1}
	& f_\text{w,scale} =\frac{w_\text{e,max,ref}}{w_\text{e,max,new}} \ \Longrightarrow \ w_\text{e,new} =\frac{w_\text{e,ref}}{f_\text{w,scale}} 
	\end{flalign}
\label{eq:scaling_3}  
\end{subequations} 
Furthermore, the new engine efficiency $\eta _\text{e,new}$ can be determined with
\begin{equation}
\label{eq:scaling_2}
\eta _\text{e,new} =\frac{T_\text{e,new} w_\text{e,new}}{\dot{m}_\text{e,ref}(w,T) \ H_\text{u}}
\end{equation}
where $H_\text{u}$ is the lower heating value of the diesel fuel, $T_\text{e,new}$ and $w_\text{e,new}$ are the scaled torque and angular velocity values obtained from (\ref{eq:scaling}) and (\ref{eq:scaling_1}) respectively. Here, the fuel flow rate map $\dot{m}_\text{e,ref}(w,T)$ of the reference engine is used. Furthermore, the engine power map is calculated by
\begin{equation}
\label{eq:Power_map}
P_\text{e} = \frac{F_\text{t}v_\text{h}}{\eta_\text{e,new} \eta_\text{t}}
\end{equation}
where $\eta_\text{t}$ is the transmission efficiency, $F_\text{t}$ and $v_\text{h}$ are the traction force and velocity respectively. The city bus is equipped with a 4-speed automatic transmission. The operating region for the gears is determined in this work using a rule-based approach with the primary objective to operate the engine at its best efficiency. These gears divide the power consumption map into four regions as shown in Fig.~\ref{fig:power_consumption_map_approximation}. 
\subsection{Approximation of ICE power consumption map}
\label{sec:Approximation}
\begin{figure}[h]
	\centering
	\includegraphics[width=8.4cm]{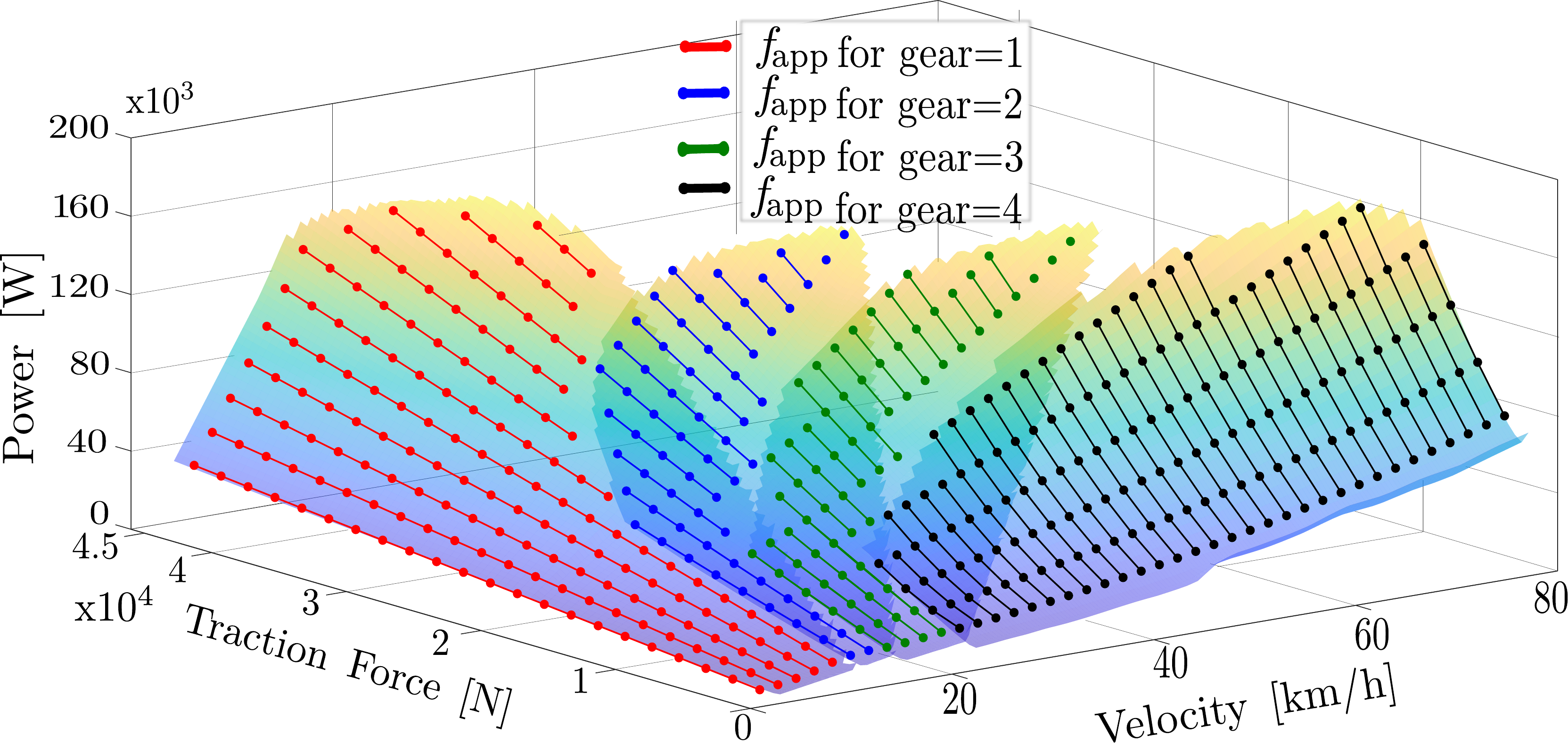}
	\caption{Power consumption map approximation}
	\label{fig:power_consumption_map_approximation}
\end{figure}

The ICE power consumption map is approximated seperately for each gear using (\ref{eq:power_consumption}) as shown in Fig.~\ref{fig:power_consumption_map_approximation}. Here, the power consumption is a function of the velocity $v_\text{h}$ and traction force $F_\text{t}$. 
\begin{equation} 
\label{eq:power_consumption}
\resizebox{0.91\hsize}{!}{$f_\text{app}(v_\text{h}, F_\text{t}) = x_{00}+x_{10}v_\text{h}+x_{01}F_\text{t} + x_{20}v_\text{h}^2+x_{02}F_\text{t}^2+x_{11}v_\text{h}F_\text{t}$}
\end{equation}
The coefficients $x_{00}$, $x_{10}$, $x_{01}$, $x_{20}$, $x_{02}$, $x_{11}$ are obtained by solving a convex optimization problem given in (\ref{eq:Power_approximation_1}). The objective is to minimize the approximation error between the polynomial function $f_\text{app}$ and the original power consumption map (\ref{eq:Power_map}) across the operating range $i=1,..,R$ and $j=1,..,S$ for the traction force and velocity respectively. Moreover, (\ref{eq:hessian_matrix}) ensures that $f_\text{app}$ is positive semidefinite. The optimization problem is solved using CVX semi-definite programming toolbox.
\begin{subequations}
	\begin{flalign} 
	\label{eq:Power_approximation_1}
	&\min_{\textit{\textbf{x}}_\text{00},\textit{\textbf{x}}_\text{10},\textit{\textbf{x}}_\text{01},\textit{\textbf{x}}_\text{20},\textit{\textbf{x}}_\text{02},\textit{\textbf{x}}_\text{11}} \sum_{i=1}^{R} \sum_{j=1}^{S} \Vert Ax-b\Vert ^{2} \\	
	&\ \ \ \ \ \ \text{s.t.}\ \ \begin{bmatrix}
	\label{eq:hessian_matrix}
	x_{02} & x_{11}\\
	x_{11} & x_{20}
	\end{bmatrix} \geqslant 0 \\
	&\text{where},\ A=\left[ 1\ F_{\text{t},i}\ v_{\text{h},j}\ F_{\text{t},i}^{2} \ v_{\text{h},j}^{2} \ F_{\text{t},i}v_{\text{h},j}\right] \nonumber \\
	&\ \ \ \ \ \ \ \ \ \ b=\frac{F_{\text{t},i}\ v_{\text{h},j}}{\eta _\text{e,new} \eta_\text{t}} \nonumber\\
	&\ \  x=[ x_{00} ,x_{01} ,x_{11} ,x_{02} ,x_{20} ,x_{11}]^{T} \nonumber	
	\end{flalign}
	\label{eq:Power_approximation_2}  
\end{subequations}  
Approximating the non-linear power consumption map using (\ref{eq:power_consumption}) has a deviation in the accuracy at lower traction values. For instance when $F_\text{t}$ is zero, the approximated power $f_\text{app}$ must ideally be zero. Thus, in a PnG glide phase while engine is OFF, the resulting energy consumption will be also zero. However, $f_\text{app}$ calculated using (\ref{eq:power_consumption}) at $F_\text{t}=0$ is found to have a value greater than zero, thus motivates the engine to remain ON at all times, which is not desirable. To tackle this issue, an idea suggested in \cite{Jia2020} is used in this work which will be explained in the next section.   
%
%
%
\section{Problem Formulations}
\label{sec:problem_formulation}
\subsection{HMPC-EACC for a vehicle-following scenario}
\label{sec:EACC_for_vehicle-following}
\begin{figure}[h]
	\centering
	\includegraphics[width=8.4cm]{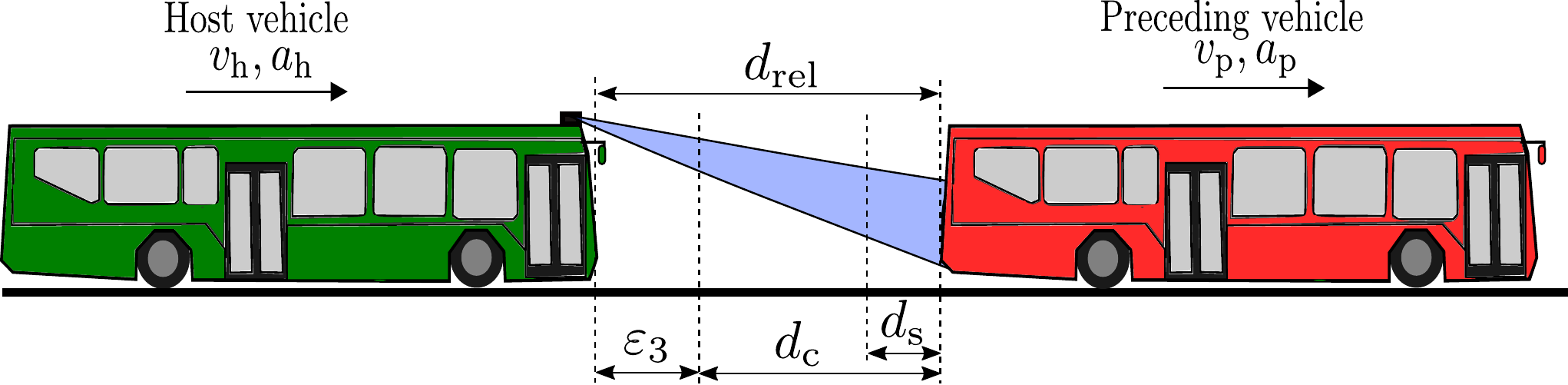}
	\caption{Typical vehicle-following schematic}
	\label{fig:initial_concept}
\end{figure}
\begin{subequations}
\begin{flalign} 
\label{eq:MPC_Cost_Function}
&\min_{\textit{P}_{\text{ICE},k},\textit{F}_{\text{t},k},\textit{F}_{\text{b},k},\textit{n}_{k},\varepsilon_{1,k},\varepsilon_{2,k},\varepsilon_{3,k}} \sum_{k=0}^{N-1} P_{\text{ICE},k}\Delta T + \zeta_{1}F_{\text{b},k}^2 + \zeta_{2}\varepsilon_{\text{1},k}^2  + \nonumber \\ &
\resizebox{0.85\hsize}{!}{$\ \ \ \ \ \ \ \ \ \ \ \ \ \ \ \ \ \ \ \ \ \ \ \ \ \ \ \ \ \ \ \ \ \ \ \ \ \ \ \ \ \zeta_{3}\varepsilon_{\text{2},k}^2 + \zeta_{4}\varepsilon_{\text{3},k}^2 $} \\ \label{eq:Online_Computation_1} 
& \resizebox{0.85\hsize}{!}{$\text{s.t.} \ \  v_{\text{h},k+1}= v_{\text{h},k} + \frac{\Delta T }{m_\text{eq}}( F_{\text{t},k}- F_{\text{b},k}  - c_\text{r}m_\text{v}g\cos\theta_{k}$} - \nonumber \\ & 
\resizebox{0.85\hsize}{!}{$\ \ \ \ \ \ \ \ \ \ \ \ \ \ \ \ \ \ \ \ \ \ \ \ \frac{1}{2}\rho A_\text{f}  c_\text{d} (p_\text{1} v_{\text{h},k}+ p_\text{2}) - m_\text{v}g\sin\theta_{k})$}  \\ \label{eq:Online_Computation_2}
& \resizebox{0.85\hsize}{!}{$d_{\text{rel},k+1}=d_{\text{rel},k}+ \Delta T \left( \frac{v_{\text{p},k}+v_{\text{p},k+1}}{2} -\frac{v_{\text{h},k}+v_{\text{h},k+1}}{2}\right)$}\\	\label{eq:Online_Computation_3} 
&\resizebox{0.45\hsize}{!}{$ \ \ \ \ \ \ \ \ \ 
	n_{k} -n_{k-1} =\ \varepsilon_{\text{1},k}$}; \ \ \ n\in \mathbb{B}:=\{0,1\}\\ \label{eq:Online_Computation_4} 
& \resizebox{0.55\hsize}{!}{$ \ \ \ \ \ \ \ \ \ \ \ \ \ \ \ \ \ \ \ \ \ 0 \leq F_{\text{t},k} \leq  F_\text{t,max} n_{k}$} \\	\label{eq:Online_Computation_5} 
&\resizebox{0.75\hsize}{!}{$  \ \ \ \ \ \ \ \ \ \ \ \
	f_\text{app}(v_{\text{h},k},F_{\text{t},k})-P_\text{max}(1-n_{k})\leq P_{\text{ICE},k}$} \\
\label{eq:Online_Computation_6}
&\resizebox{0.55\hsize}{!}{$  \ \ \ \ \ \ \ \ \ \ \ \ \ \ \ \ \ \ \ \ \ \ 
	P_{\text{ICE},k} \leq P_\text{max}  n_{k}$} \\
\label{eq:Online_Computation_7}
&\resizebox{0.6\hsize}{!}{$ \ \ \ \ \ \ \ \ \ \ \ \ \ \ \ v_{\text{min},k} \leq v_{\text{h},k} \leq  v_{\text{max},k}$} \\ \label{eq:Online_Computation_8}   
& \resizebox{0.55\hsize}{!}{$ \ \ \ \ \ \ \ \ \ \ \ \ \ \ \ \ \ \ \ \ \ \ 0 \leq F_{\text{b},k} \leq  F_\text{b,max}$} \\	\label{eq:Online_Computation_9} 
&\resizebox{0.65\hsize}{!}{$  \ \ \ \ \ \ \ \ \ \ \ \ \ \ \ \ |F_{\text{t},k}- F_{\text{t},k+1}| - \varepsilon_{\text{2},k} \leq \Delta F_\text{t,max}$}  \\ \label{eq:Online_Computation_10}
& \ \ \ \ \ \ \ \ \ \ \ \ \ \ \ d_{\text{rel},k} \geq d_\text{min} + h_\text{s} v_{\text{h},k} \\ \label{eq:Online_Computation_11} 
& \ \ \ \ \ \ \ \ \ \  d_{\text{rel},k} \leq d_\text{max} + h_\text{c} v_{\text{h},k} + \varepsilon_{\text{3},k}
	\end{flalign}
	\label{eq:Online_Computation_12}  
\end{subequations} 
The HMPC-EACC optimization problem for a typical vehicle-following scenario is reformulated  in time-domain based on the idea from \cite{Jia2020} and is given in (\ref{eq:MPC_Cost_Function}). Here, $N$ is the prediction horizon length. The first term in the cost function represents the energy consumption of the host vehicle which is the product of the power consumption $P_\text{ICE}$ and the sample time $\Delta T$. The variable $P_\text{ICE}$ is constrained by (\ref{eq:Online_Computation_5}) and (\ref{eq:Online_Computation_6}). In order to compensate the error in power consumption approximation as discussed in Section \ref{sec:Approximation}, an additional term $P_\text{max}(1-n_{k})$ is subtracted from $f_\text{app}$ in (\ref{eq:Online_Computation_5}), where $P_\text{max}$ is a constant term, which is the maximum available power at each gear and is computed offline from the original map and $n_{k}$ is given by
\begin{equation}
\label{eq:Engine_state}
n_{k} =\begin{cases}
0 & : \text{Engine OFF}\\
1 & : \text{Engine ON}
\end{cases}
\end{equation}
If the engine is OFF, both (\ref{eq:Online_Computation_5}) and (\ref{eq:Online_Computation_6}) force $P_\text{ICE}$ to zero. Expanding $f_\text{app}$ term in (\ref{eq:Online_Computation_5}) by substituting (\ref{eq:power_consumption}), makes it a quadratically constrained problem. Using the second term in (\ref{eq:MPC_Cost_Function}), an excessive braking force $F_\text{b}$ is penalized. In the third term, a slack variable $\varepsilon_{1}$ is introduced to avoid frequent switching between the engine ON and OFF states. A linear constraint describing $\varepsilon_{1}$ is given in (\ref{eq:Online_Computation_3}), where $n_{k}$ is the current and $n_{k-1}$ is the previous engine state. Moreover, the variable $n$ takes a binary value either 0 or 1 as described in (\ref{eq:Engine_state}). Furthermore, in the fourth term the variation in the traction force at successive time steps is penalized using a slack variable $\varepsilon_{2}$ if it exceeds a constant value $\Delta F_\text{t,max}$, with the aim to minimize the jerks and improve the driving comfort, as given in (\ref{eq:Online_Computation_9}). To motivate the host vehicle to stay within the desired region $d_\text{c}$, another slack variable $\varepsilon_{3}$ is penalized in the last term. The host vehicle velocity and relative distance to the leading vehicle in discretized form is given in (\ref{eq:Online_Computation_1}) and (\ref{eq:Online_Computation_2}), where $v_\text{p}$ is the velocity of the preceding vehicle. Moreover, $\zeta_{1}$, $\zeta_{2}$, $\zeta_{3}$, $\zeta_{4}$ are the corresponding weighting factors for the aforementioned terms in the cost function (\ref{eq:MPC_Cost_Function}). The limits for the traction force $F_\text{t}$ are set using (\ref{eq:Online_Computation_4}), in which $F_\text{t,max}$ is the maximum traction force calculated offline for each gear. Moreover, $n_{k}$ is multiplied to $F_\text{t,max}$ so as to force $F_\text{t}$ to zero when the engine is OFF. The physical limitations of the vehicle with respect to velocity and braking force are addressed in (\ref{eq:Online_Computation_7}) and (\ref{eq:Online_Computation_8}). In order to maintain a safe distance to the preceding vehicle, a hard constraint is introduced in (\ref{eq:Online_Computation_10}) based on (\ref{eqn:SystemDyn_Cont_4}). To motivate the host vehicle to stay within the desired region a soft constraint is formulated using (\ref{eq:Online_Computation_11}) based on (\ref{eqn:SystemDyn_Cont_5}). Overall, the presence of a quadratic cost function (\ref{eq:MPC_Cost_Function}), a binary variable (\ref{eq:Online_Computation_3}), linear (\ref{eq:Online_Computation_4}), (\ref{eq:Online_Computation_6}), (\ref{eq:Online_Computation_7}), (\ref{eq:Online_Computation_8}), (\ref{eq:Online_Computation_9}), (\ref{eq:Online_Computation_10}), (\ref{eq:Online_Computation_11}) and quadratic inequality constraints (\ref{eq:Online_Computation_5}) make the optimization problem as MIQCQP, which can be solved efficiently using Gurobi solver. 
\subsection{HMPC-EACC for a typical urban scenario}
	\label{sec:HMPC-EACC_urban}
\begin{figure}[h]
	\centering
	\includegraphics[width=8.4cm]{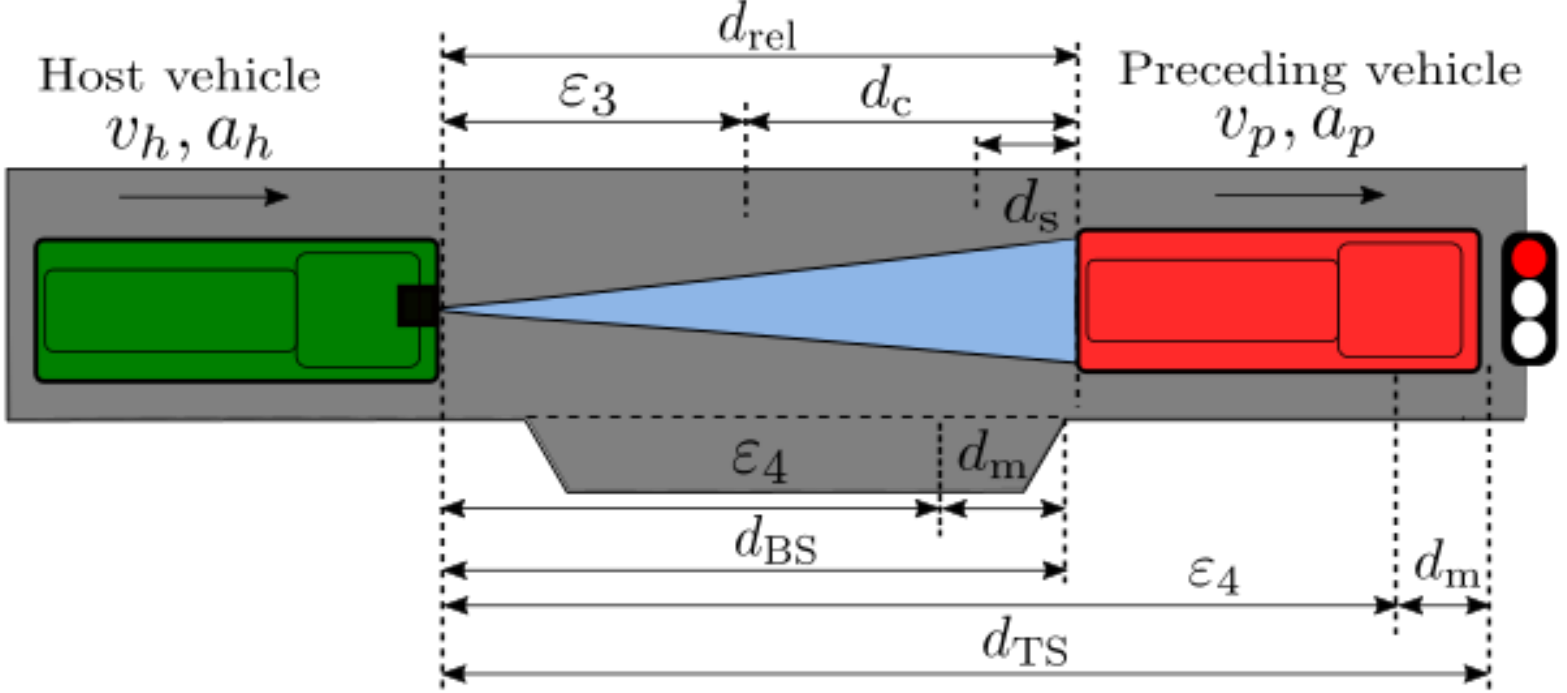}
	\caption{Typical urban scenario schematic}
	\label{fig:Car_followingschematic}
\end{figure}
The previous optimization problem (\ref{eq:MPC_Cost_Function}) is extended in this section to accomodate the additional constraints due to TL signals and bus stops during the longitudinal control. The new cost function is presented in (\ref{eq:MPC_Cost_Function_2}), which has two new terms. The penultimate term contains the slack variable $\varepsilon_{4}$ to motivate the host vehicle to stop as close as possible to the stop line $d_\text{m}$ while stopping at a TL or bus stop as defined in (\ref{eq:Online_Computation_16}). Here, $d_\text{ITS}$ is the relative distance to the upcoming stop that could be either a TL signal or a bus stop, where $\text{ITS}=\{\text{BS},\text{TS}\}$.
\begin{subequations}
	\begin{flalign} 
\label{eq:MPC_Cost_Function_2}
&\ \ \ \ \ \min_{u_{k}} \sum_{k=0}^{N-1} P_{\text{ICE},k}\Delta T + \zeta_{1}F_{\text{b},k}^2  + \zeta_{2}\varepsilon_{\text{1},k}^2 + \zeta_{3}\varepsilon_{\text{2},k}^2 +   \nonumber \\ &
\resizebox{0.85\hsize}{!}{$  \ \ \ \ \ \ \ \ \ \ \ \ \ \ \ \ + \zeta_{4}\varepsilon_{\text{3},k}^2 + \zeta_{5}\varepsilon_{\text{4},k}^2 + \zeta_{6}(v_{\text{h},k}-v_{\text{ref},k})^2 $} \\ \label{eq:Online_Computation_13} 
	&  \ \ \ \ \ \ \ \  \text{s.t.} \ \ \ \    \ \ \ \ 	\resizebox{0.4\hsize}{!}{$  x_{k+1} =Ax_{k} +Bu_{k} +w_{i}$} \\ \label{eq:Online_Computation_14} 
	& \resizebox{0.95\hsize}{!}{$  \ \ \ \  \ \ \ \ \  (\text{\ref{eq:Online_Computation_3}}), (\text{\ref{eq:Online_Computation_4}}), (\text{\ref{eq:Online_Computation_5}}), (\text{\ref{eq:Online_Computation_6}}), (\text{\ref{eq:Online_Computation_7}}), (\text{\ref{eq:Online_Computation_8}}), (\text{\ref{eq:Online_Computation_9}}), (\text{\ref{eq:Online_Computation_10}}), (\text{\ref{eq:Online_Computation_11}})$} \nonumber \\
	& \resizebox{0.6\hsize}{!}{$  \ \ \ \ \ \ \ \ \ \ \ \ \ \ \ \ \ \ \ \ \ \ \ \ \ \ \ \ \ \ \ d_{\text{ITS},k} \geq 0$} \\	\label{eq:Online_Computation_16}   	
	&\resizebox{0.65\hsize}{!}{$   \ \ \ \ \ \ \ \ \ \ \ \ \ \ \ \ \ \ \ \ \ \ \ \ \ \  d_{\text{ITS},k} - \varepsilon_{\text{4},k} \leq d_\text{m}$}
	\end{flalign}
	\label{eq:Online_Computation_24}  
\end{subequations}
where
\begin{subequations}
	\begin{flalign} 
	&\resizebox{0.95\hsize}{!}{$A = \begin{bmatrix}
		A_\text{11} & 0 & 0\\
		-A_\text{21} & 1 & 0\\
		-\Delta T & 0 & 1
		\end{bmatrix} \ ,\ B=\begin{bmatrix}
		B_\text{11} & -B_\text{12} & 0 & \cdots  & 0\\
		-B_\text{21} & B_\text{22} & 0 & \cdots  & 0\\
		0 & 0 & 0 & \cdots  & 0
		\end{bmatrix} ,\ w=\begin{bmatrix}
		-w_{\text{1},k}\\
		w_{\text{2},k}\\
		0
		\end{bmatrix}$}\nonumber \\
	& \ \ \ \ \ \ \ \ \ \ \ \ \ \resizebox{0.65\hsize}{!}{$A_\text{11} =\left( 1-\frac{\Delta T}{m_\text{eq}} \psi\right) ,\ A_\text{21} =\frac{\Delta T}{2}\left( 2-\frac{\Delta T}{m_\text{eq}} \psi\right) ,$}\nonumber \\
	& \ \ \ \ \ \ \ \ \ \ \ \ \  \resizebox{0.55\hsize}{!}{$\ B_\text{11} =B_\text{12} =\frac{\Delta T}{m_\text{eq}} ,\ B_\text{21} =B_\text{22} =\frac{\Delta T^{2}}{2m_\text{eq}} , $}\nonumber \\
	& \ \ \ \ \  \resizebox{0.85\hsize}{!}{$w_{\text{1},k} =\frac{\Delta T}{m_\text{eq}} \ ( m_\text{v}g\sin\theta_{k} +c_\text{r}m_\text{v}g \cos\theta_{k} + \frac{1}{2} \rho A_\text{f}c_\text{a}p_\text{1}), \ $} \nonumber \\
	& \ \ \ \ \ \ \ \ \ \ \ \ \ \resizebox{0.6\hsize}{!}{$w_{\text{2},k} =\frac{\Delta T}{2}( v_{\text{p},k} +v_{\text{p},k+1} -w_{1,k}),$} \nonumber \\
	& \ \ \ \ \ \ \ \ \ \ \ \ \  \ \ \ \ \ \ \ \ \ \ \ \  \resizebox{0.25\hsize}{!}{$\psi =\frac{1}{2} \rho A_\text{f} c_\text{a} p_\text{2}$}\nonumber 
	\end{flalign}
	\label{eq:Online_Computation_26}  
\end{subequations}
 The last term in (\ref{eq:MPC_Cost_Function_2}) enforces a penalty for the deviation of the host vehicle velocity $v_\text{h}$ from the GWOS reference velocity $v_\text{ref}$ calculated from (\ref{eqn:intersection}). The inequality constraint (\ref{eq:Online_Computation_14}) is enforced so that the host vehicle does not cross TL signals when the phase is red. The other constraints remain the same as in (\ref{eq:Online_Computation_12}). The discrete state space representation of system dynamics in time-domain is given by (\ref{eq:Online_Computation_13}). Here, the state variables are $x_{k}= [v_{\text{h},k}, d_{\text{rel},k}, d_{\text{ITS},k}]^\intercal$ and the optimization variables are $u_{k}= [P_{\text{ICE},k},F_{\text{t},k},F_{\text{b},k},n_{k},\varepsilon_{\text{1},k},\varepsilon_{\text{2},k}, \varepsilon_{\text{3},k}, \varepsilon_{\text{4},k}]^\intercal$.  It is important to note that, $\zeta_{5}$ is initially zero and is activated when approaching a red traffic light. Moreover, in a vehicle-following scenario, $\zeta_{6}$ is deactivated and in a freeway scenario, $\zeta_{4}$ is deactivated by choosing them as zero. 
 
\subsection{Baseline controller for a typical urban scenario}
	\label{sec:Baseline_controller}	
To validate the performance of the HMPC-EACC controller proposed in the previous section, a problem formulation for the baseline controller is chosen from \cite{Jia2018}. Although this problem formulation did not consider the influence of the TL signals on the host vehicle during car-following, it has been improvised in this section to have a fair comparison with HMPC-EACC. The new problem formulation is given in (\ref{eq:MPC_Cost_Function_3}).
\begin{subequations}
	\begin{flalign} 
	\label{eq:MPC_Cost_Function_3}
	&\min_{\textit{F}_{\text{t},k},\textit{F}_{\text{b},k},\varepsilon_{1,k},\varepsilon_{2,k},\varepsilon_{3,k}} \sum_{k=0}^{N-1} f_\text{app}(v_{\text{h},k}, F_{\text{t},k}) + \zeta_{1}F_{\text{b},k}^2 + \zeta_{2}\varepsilon_{\text{2},k}^2  + \nonumber \\ &
	\resizebox{0.85\hsize}{!}{$\ \ \ \ \ \ \ \ \ \ \ \ \ \ \ \ \ \ \ \ \ \ \ \ \ \ \ \ \ \ \ \ \ \ \ \ \ \ \ \ \ \zeta_{3}\varepsilon_{\text{3},k}^2 + \zeta_{4}\varepsilon_{\text{4},k}^2 $} \\
	& \ \ \ \ \text{s.t.} \resizebox{0.75\hsize}{!}{$  \ \ \ \  (\text{\ref{eq:Online_Computation_13}}),(\text{\ref{eq:Online_Computation_14}}), (\text{\ref{eq:Online_Computation_16}}),(\text{\ref{eq:Online_Computation_7}}),(\text{\ref{eq:Online_Computation_8}}),(\text{\ref{eq:Online_Computation_9}}),(\text{\ref{eq:Online_Computation_10}}),(\text{\ref{eq:Online_Computation_11}}) $} \nonumber \\
	\label{eq:Online_Computation_17}   	
	&\resizebox{0.65\hsize}{!}{$ \ \ \ \ \ \ \ \ \ \ \ \ \ \ \ \ \ \ \ \ \ \ \ \ \ \ \ \ 0 \leq F_{\text{t},k} \leq  F_\text{t,max}$}
	\end{flalign}
	\label{eq:Online_Computation_18}  
\end{subequations} 
In comparison to the HMPC-EACC problem formulation (\ref{eq:MPC_Cost_Function_2}), one major difference in (\ref{eq:MPC_Cost_Function_3}) is that the first term represents the approximated power consumption obtained from (\ref{eq:power_consumption}), meaning that the PnG strategy cannot be realized in the baseline controller. Therefore, as there is no necessity for the slack variable $\varepsilon_{1}$, it is eliminated in (\ref{eq:MPC_Cost_Function_3}). Another major difference is that the baseline controller does not have the possibility to track a reference velocity $v_\text{ref}$ for green-wave and performs vehicle-following throughout the trip. Therefore, the reference tracking term is eliminated in (\ref{eq:MPC_Cost_Function_3}). The additional constraints due to TL signals and bus stops are added to (\ref{eq:Online_Computation_18}) as defined in the previous problem formulation.
\section{Results}
\label{sec:results}
\subsection{Vehicle-following scenario}
	\label{sec:Result_Vehicle-following_scenario}
\begin{figure}[t]
	\centering
	\includegraphics[width=8.4cm]{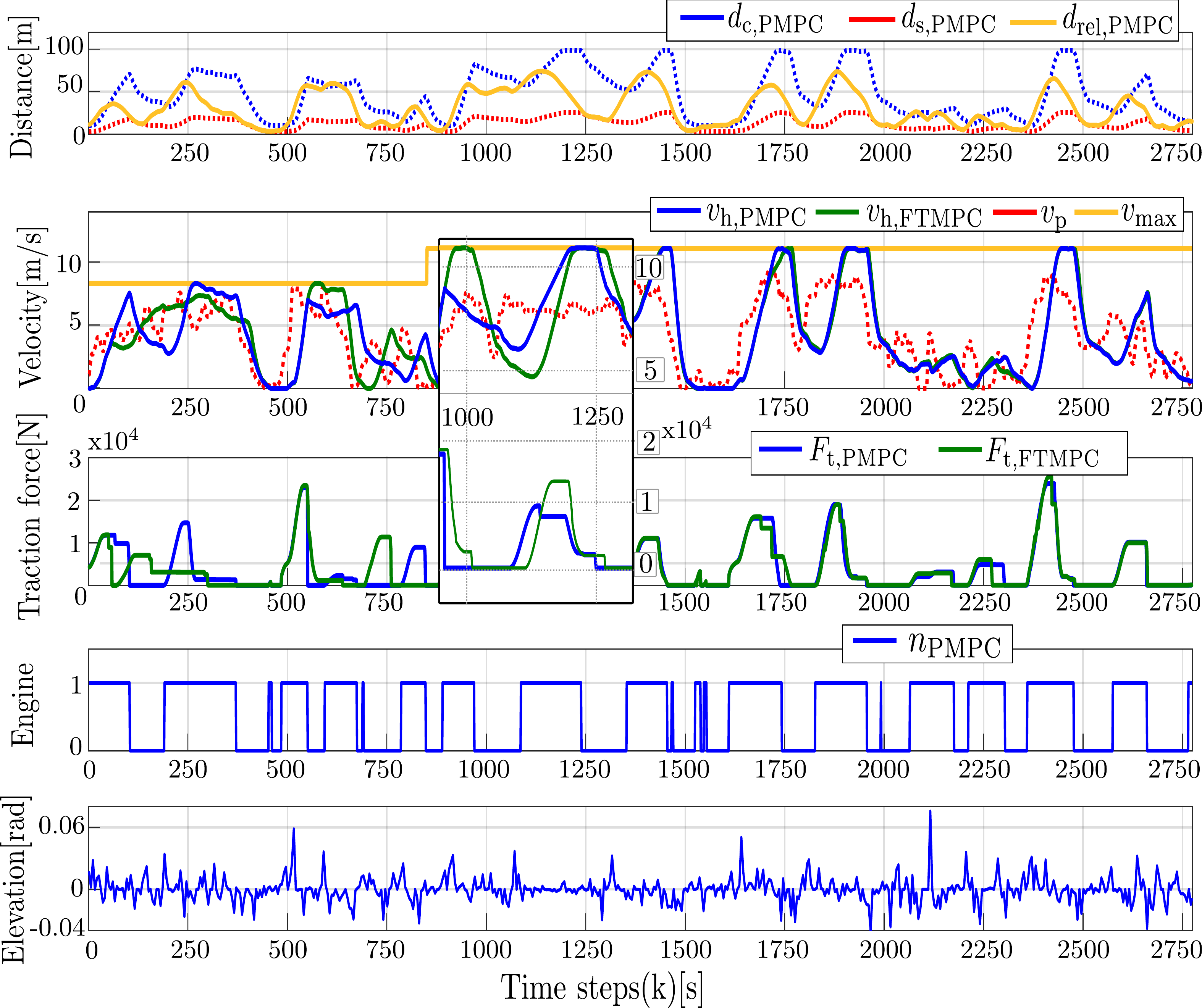}
	\caption{Results of HMPC-EACC for a vehicle-following scenario}
	\label{fig:Car_following_result}
\end{figure}

To demonstrate the performance of the HMPC-EACC problem formulation discussed in Section \ref{sec:EACC_for_vehicle-following}, it is tested on a city bus velocity profile gathered from real-world tests. In this simulation study, this velocity profile is considered as the actual speed profile of a leading vehicle and is tracked by the proposed controller. Two deterministic control approaches are chosen for this investigation, namely frozen-time model predictive control (FTMPC) and prescient model predictive control (PMPC). FTMPC has the knowledge of only the current velocity information of the preceding vehicle and assumes the velocity to be constant throughout the prediction horizon, whereas PMPC assumes that the future velocity information is perfectly known within the prediction horizon. The results for a typical vehicle-following scenario are illustrated in Fig.~\ref{fig:Car_following_result} and it is evident that the host vehicle equipped with PMPC and FTMPC have smoother velocity profiles $v_\text{h,PMPC}$ and $v_\text{h,FTMPC}$ respectively, as compared to the preceding vehicle $v_\text{p}$ while performing vehicle-following. Moreover, the host vehicle strictly adheres to the regulatory road speed limits $v_\text{max}$. The host vehicle is seen to robustly maintain the inter-vehicle distance $d_\text{rel}$ and resides within the desired region $d_\text{c}$ without crossing the safe distance $d_\text{s}$ to the preceding vehicle.
Furthermore, the host vehicle applies the PnG strategy at several time intervals, resulting in switching the engine OFF and ensures that the traction force is zero, for example at \unit[104]{s}, \unit[375]{s}, \unit[1760]{s} and \unit[2500]{s}.
\begin{table}[t]
		\begin{center}
	\caption{Simulation results}
	\begin{tabular}{cccc}
			\hline\hline
		\begin{tabular}[c]{@{}c@{}}Prediction\\ Horizon\end{tabular} & Controller        & \begin{tabular}[c]{@{}c@{}}Distance\\ {[}km{]}\end{tabular} & \begin{tabular}[c]{@{}c@{}}Fuel\\ savings\end{tabular} \\ \hline
		\multirow{3}{*}{N=8}                                         & Preceding vehicle & 2.582                                                       & -                                                      \\
		& HMPC-EACC (FTMPC) & 2.578                                                       & +7.11\%                                                \\
		& HMPC-EACC (PMPC)  & 2.578                                                       & +12.11\%                     \\ \hline                                
	\end{tabular}
	\label{tab:ECO-MPC1_extension}
	\end{center}
\end{table} 
To demonstrate the fuel consumption for the proposed controllers, a longitudinal vehicle model for the city bus has been developed using QSS toolbox. The results for the aforementioned scenario are summarized in Table~\ref{tab:ECO-MPC1_extension}. It can be noticed that the HMPC-EACC host car equipped with PMPC and FTMPC can achieve up to 12.11\% and 7.11\% reductions in fuel consumption as compared to the preceding vehicle. The reason is that the HMPC-EACC uses the future information of the route elevation and the preceding vehicle future velocities to derive optimal control inputs for the host vehicle. Moreover, gliding by switching the engine OFF using the PnG strategy is another important reason for obtaining better energy savings. The reason for higher energy savings with PMPC as compared to FTMPC is primarily due to the availability of perfect knowledge of the preceding vehicle's future velocities. It can be noticed from the enlarged view in Fig.~\ref{fig:Car_following_result}, that the host vehicle velocity with FTMPC has bigger fluctuations while tracking the preceding vehicle as compared to the PMPC and corresponding traction force curves show higher peaks for FTMPC as compared to PMPC. 
\subsection{Evaluation of HMPC-EACC and baseline controller}
 \begin{figure}[t]
	\centering
	\includegraphics[width=8.4cm]{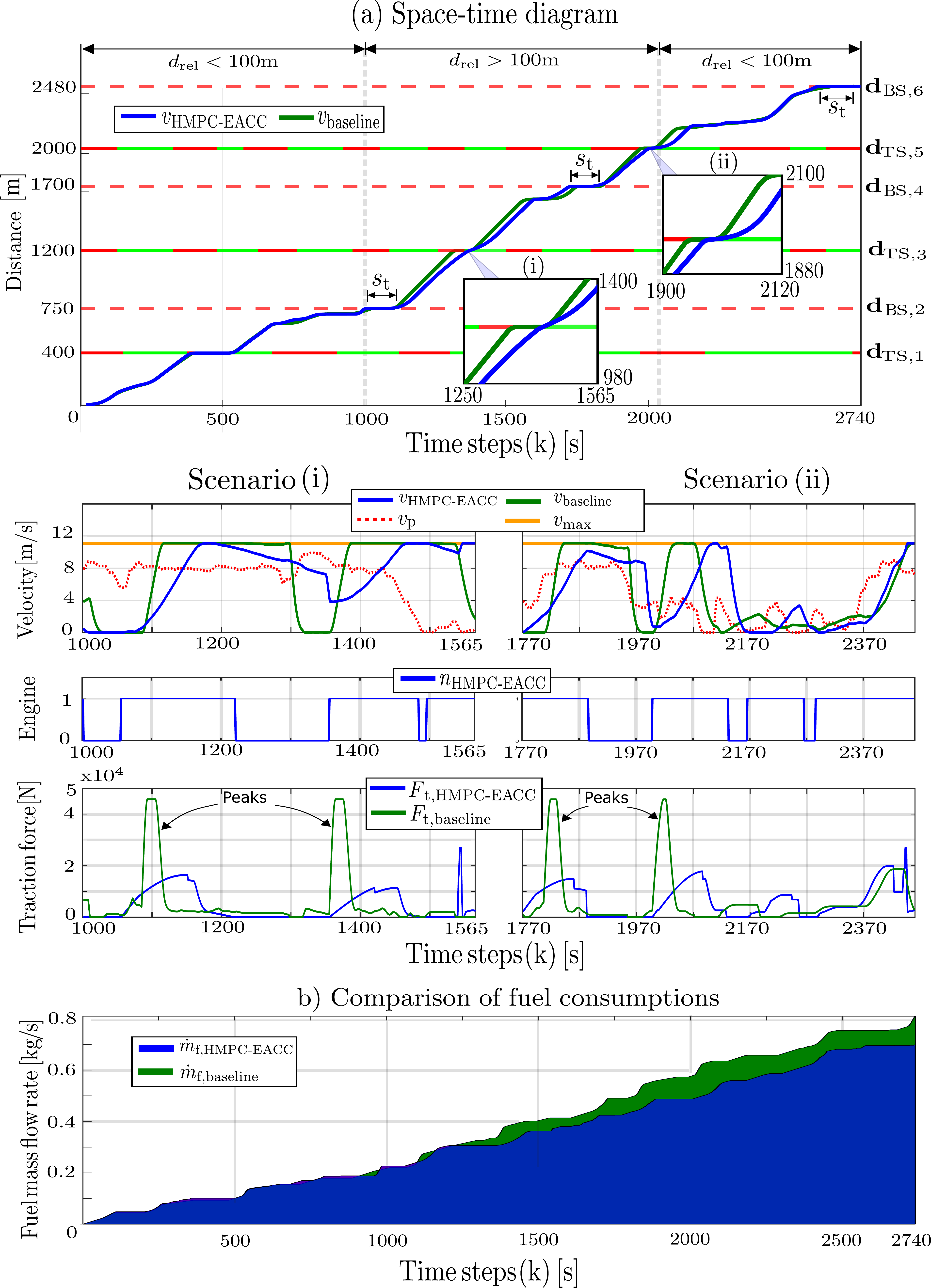}
	\caption{Comparison of the HMPC-EACC and baseline controller}
	\label{fig:Car_following_result_2}
\end{figure}
\begin{table}[t]
	\caption{Simulation results}
	\begin{center}
		\begin{tabular}{ccccc}
			\hline\hline
			\begin{tabular}[c]{@{}c@{}}Prediction\\ Horizon\end{tabular} & Controller & \begin{tabular}[c]{@{}c@{}}Distance\\ {[}km{]}\end{tabular} & \begin{tabular}[c]{@{}c@{}}Trip time\\ {[}s{]}\end{tabular} & \begin{tabular}[c]{@{}c@{}}Fuel\\ savings\end{tabular} \\ \hline
			\multirow{2}{*}{$N$=8}                                         & Baseline   & 2.48                                                        & 2739                                                        & -                                                      \\
			& HMPC-EACC  & 2.48                                                        & 2737                                                        & +10.21\%                                               \\ \hline
			\multirow{2}{*}{$N$=15}                                        & Baseline   & 2.48                                                        & 2740                                                        & -                                                      \\
			& HMPC-EACC  & 2.48                                                        & 2735                                                        & +12\%                                               
			\\ \hline \hline                                        
		\end{tabular}
	\end{center}
	\label{tab:ECO-MPC1_extension_2}
\end{table}
In this section, the performance of the proposed HMPC-EACC and baseline controller from Section \ref{sec:HMPC-EACC_urban} and \ref{sec:Baseline_controller} respectively, is evaluated in a typical urban scenario in the presence of a preceding vehicle, signalized intersections and bus stops. Moreover, it is assumed that both the controllers have the perfect knowledge of the preceding vehicle future velocities and location of the upcoming bus stops. However, the major difference is that the host vehicle with HMPC-EACC receives the SPaT information of the upcoming TL signals in real-time, whereas the baseline controller receives only the current phase information of the TL signals.  
Both the controllers are tested in a simulative urban driving route of approximately \unit[2.5]{km} stretch with signalized intersections $\text{TS}_{\{1,3,5\}}$ and bus stops $\text{BS}_{\{2,4,6\}}$ as illustrated in the space-time diagram in  Fig.~\ref{fig:Car_following_result_2}a.
In this simulative study, the HMPC-EACC initially begins by tracking the same preceding vehicle velocity profile introduced in Section \ref{sec:Result_Vehicle-following_scenario}. At the first signalized intersection $\text{TS}_{1}$ around \unit[460]{s}, the host vehicle decelerates to stop behind the preceding vehicle because of a red TL phase. Later at \unit[1000]{s}, the host vehicle is seen to make the compulsory stop at the bus stop $\text{BS}_{2}$ for a constant dwell time $s_\text{t}$, therefore the relative distance $d_\text{rel}$ between the host vehicle and preceding vehicle becomes larger. As soon as $d_\text{rel}>$\unit[100]{m} (outside sensor range) the HMPC-EACC stops tracking the preceding vehicle and follows a GWOS to reach the upcoming TL signals at a green phase. Moreover, two scenarios encountered at the signalized intersections $\text{TS}_{\{3,5\}}$ are enlarged and depicted in Fig.~\ref{fig:Car_following_result_2}a. A more detailed analysis of these scenarios is presented in Fig.~\ref{fig:Car_following_result_2}(i) and Fig.~\ref{fig:Car_following_result_2}(ii) respectively. At \unit[1322]{s} and \unit[1986]{s},
the baseline vehicle decelerates to stop at these TL signals, due to the fact that the baseline controller lacks the SPaT information of the immediate TL signals. On the other hand, the host vehicle equipped with HMPC-EACC has the perfect knowledge of SPaT and crosses the TL signals at a green phase. Moreover, as compared to the host vehicle traction force $F_\text{t,HMPC-EACC}$, it can be noticed that the traction force values for the baseline vehicle $F_\text{t,baseline}$ are found to be very large, as a result of the strong deceleration and acceleration maneuvers incurred while stopping at the TL signals. Furthermore, the fuel mass flow rate of both the host and baseline vehicles for the entire simulation length is plotted in Fig.~\ref{fig:Car_following_result_2}b. It is evident that the baseline vehicle has consumed more fuel in comparison with the HMPC-EACC, because of the aforementioned reasons. The overall test results are summarized in Table~\ref{tab:ECO-MPC1_extension_2}. The host vehicle equipped with the proposed HMPC-EACC has outperformed the baseline controller by achieving fuel savings up to 10.2\% and 12\% for a prediction horizon length of $N$=~\unit[8] and $N$=~\unit[15] respectively in a realistic urban scenario. 
\section{Computation Time}
	\label{sec:computation_time}
The execution time for the proposed HMPC-EACC strategy at different prediction horizons is summarized in Table \ref{tab:Execution_time_2}. The evaluation is performed using Matlab$\textsuperscript{\textregistered}$ R2020a profiler on a Windows 10 PC equipped with an Intel$\textsuperscript{\textregistered}$ Core$\textsuperscript{\texttrademark}$ i7-7500U CPU processor with 2.70 GHz clock frequency and 12 GB RAM. For a predition horizon of $N$=~\unit[8], $N$=~\unit[15] and $N$=~\unit[25], the mean execution time for the proposed controller is found to be \unit[23]{ms}, \unit[71]{ms} and \unit[224]{ms} respectively. The sample time is chosen as $\Delta T$=\unit[200]{ms} in this work and therefore the results indicate that the prediction horizon length must be less than $N$=~\unit[25] steps (\unit[5]{s}) to realize online-implementation. 
\begin{table}[h]
	\begin{center}		
		\caption{Computation time for different prediction horizons}\label{real_time}
		\begin{tabular}{c|cccccc}
			\hline \hline
			Prediction horizon & $N$=8 & $N$=15 & $N$=25 \\\hline
			Computation time & \unit[23]{ms} & \unit[71]{ms} & \unit[224]{ms} \\\hline
		\end{tabular}
	\label{tab:Execution_time_2}
	\end{center}
\end{table}

\section{CONCLUSIONS}
\label{sec:conclusions}
In this work, with the goal to minimize the fuel consumption in a city bus with ICE, an ecological ACC concept based on HMPC is proposed. Firstly, a problem formulation for HMPC-EACC is designed in time-domain for a typical vehicle-following scenario and is evaluated on a real-world city bus profile. The results revealed that the host vehicle can achieve fuel savings up to 12.1\% while tracking a leading vehicle by implementing the PnG strategy. Furthermore, a new problem formulation for HMPC-EACC is proposed that can handle additional constraints due to TL signals and bus stops in an urban scenario, besides tracking a preceding vehicle. The proposed controller uses the SPaT information to devise a GWOS for the host vehicle, to avoid unnecessary stops at the TL signals, which has shown additional fuel saving benefits up to 12\% when tested against a baseline controller. Furthermore, a deeper analysis on the computation time for different prediction horizons is made. The results showcased that the proposed controller has the capability for online-implementation. In future work, the proposed HMPC-EACC concept will be evaluated on more driving datasets, in the presence of both deterministic and actuated traffic light signals. Furthermore, an extension of EACC with a possibility for lateral control, which is a limitation in the current work will be investigated in the future work.
\section*{Acknowledgment}

This work is funded by the German Ministry for Education and Research (BMBF) and partly supported by the Center for Commercial Vehicle Technology (ZNT) at the University of Kaiserslautern, Germany.

\bibliographystyle{IEEEtran}
\bibliography{IEEE2021}

\end{document}